\documentclass[5p,times]{elsarticle} 

\usepackage{amssymb}
\usepackage{amsmath}
\usepackage{lineno} 
\usepackage[T1]{fontenc}

\journal{Nuclear Inst. and Methods in Physics Research, A}

\begin{document}

\begin{frontmatter}

\title{The xpDIRC Concept for Next-Generation DIRC Detectors}

\author[3] {R.~Dzhygadlo\corref{cor1}}
\ead{r.dzhygadlo@gsi.de}
\cortext[cor1]{Corresponding author}

\author[1] {J.~Datta} 
\author[10] {K.~Dehmelt} 
\author[1,9]{A.~Deshpande}
\author[1] {T.~K.~Hemmick}
\author[4] {Md. I.~Hossain}
\author[5] {C.E.~Hyde}
\author[6] {Y.~Ilieva}
\author[4] {G.~Kalicy}
\author[7] {W.J.~Llope}
\author[2,6] {P.~Nadel-Turonski}
\author[3] {C.~Schwarz}
\author[3] {J.~Schwiening}
\author[1] {N.~Shankman}
\author[8] {J.~Stevens}
\author[4] {N.~Wickramaarachchi}
\author[9] {C.~Woody}

\affiliation[3]{organization={GSI Helmholtzzentrum f\"ur Schwerionenforschung GmbH},
  city={Darmstadt},
  postcode={64291},
  country={Germany}}
\affiliation[1]{organization={Department of Physics, Stony Brook University},
  city={Stony Brook},
  postcode={NY 11794},
  country={USA}}
\affiliation[2]{organization={Center for Frontiers in Nuclear Science, Stony Brook University},
  city={Stony Brook},
  postcode={NY 11794},
  country={USA}}
\affiliation[4]{organization={Catholic University of America},
  city={Washington},
  postcode={DC 20064},
  country={USA}}
\affiliation[5]{organization={Old Dominion University},
  city={Norfolk},
  postcode={VA 23529},
  country={USA}}
\affiliation[6]{organization={University of South Carolina},
  city={Columbia},
  postcode={SC 29208},
  country={USA}}
\affiliation[7]{organization={Wayne State University},
  city={Detroit},
  postcode={MI 48201},
  country={USA}}
\affiliation[8]{organization={College of William \& Mary},
  city={Williamsburg},
  postcode={VA 23185},
  country={USA}}
\affiliation[9]{organization={Brookhaven National Lab},
  city={Upton},
  postcode={NY 11973},
  country={USA}}
\affiliation[10]{organization={Thomas Jefferson National Accelerator Facility},
  city={Newport News},
  postcode={VA 23606},
  country={USA}}

\begin{abstract}

The next-generation DIRC (xpDIRC) represents a novel detector geometry concept
currently being developed for advanced particle identification systems
in high-energy physics experiments.
Building upon the high-performance DIRC (hpDIRC) designed
for the ePIC detector at the Electron-Ion Collider,
the xpDIRC introduces a hybrid optical architecture that
combines enhanced focusing optics, a wide plate light guide,
and a compact expansion volume.
Comprehensive Geant4 simulations demonstrate that the xpDIRC
hybrid geometry achieves state-of-the-art performance in $\pi/K$
separation across the entire range of operation.

\end{abstract}


\end{frontmatter}


\section{Introduction}

The DIRC (Detection of Internally Reflected Cherenkov light)
technology offers a compact and effective solution for
charged particle identification (PID) in modern collider experiments.
From its first application in BaBar \cite{babar}
to its more recent versions in Belle II \cite{belle2top}
and GlueX \cite{gluex}, the technology has reliably delivered
$\pi/K$ separation at momenta up to 3.5 GeV/c.
The ePIC high-performance DIRC (hpDIRC) \cite{hpdirc} now
extends this performance to 6 GeV/c, using small-pixel
MCP-PMT sensors, a novel 3-layer spherical lens,
and fast readout electronics.
The xpDIRC concept builds on this foundation by
exploring the physical performance limits of DIRC technology.
The outcomes of these studies would
be applicable to both ePIC and, in particular,
to Detector-2 \cite{det2,det2b} at the Electron-Ion Collider \cite{eic}.

\section{Detector design and performance}

Following the design of the ePIC hpDIRC, the xpDIRC is envisioned
as a barrel-type detector composed of isolated sectors.
Each sector comprises 10 narrow radiator bars of $17\times35\times3600$~mm$^3$ size, a focusing system,
a wide plate light guide of $17\times350\times840$~mm$^3$ size, and a compact, prism-shaped expansion volume (EV)
equipped with an array of photosensors.
The use of a wide plate as a light guide in DIRC technology
was first proposed as a natural evolution of the fDIRC concept \cite{fdirc}.
The focusing system is implemented using a 3-layer spherical lens attached to each radiator bar,
a configuration first introduced and tested for the PANDA Barrel DIRC \cite{pandadirc}.
Combining this type of focusing with a plate light guide
enhances performance while reducing the overall cost
and, potentially, the size of the readout section.

\begin{figure}[ht]
  \centering
  \includegraphics[width=0.9\columnwidth]{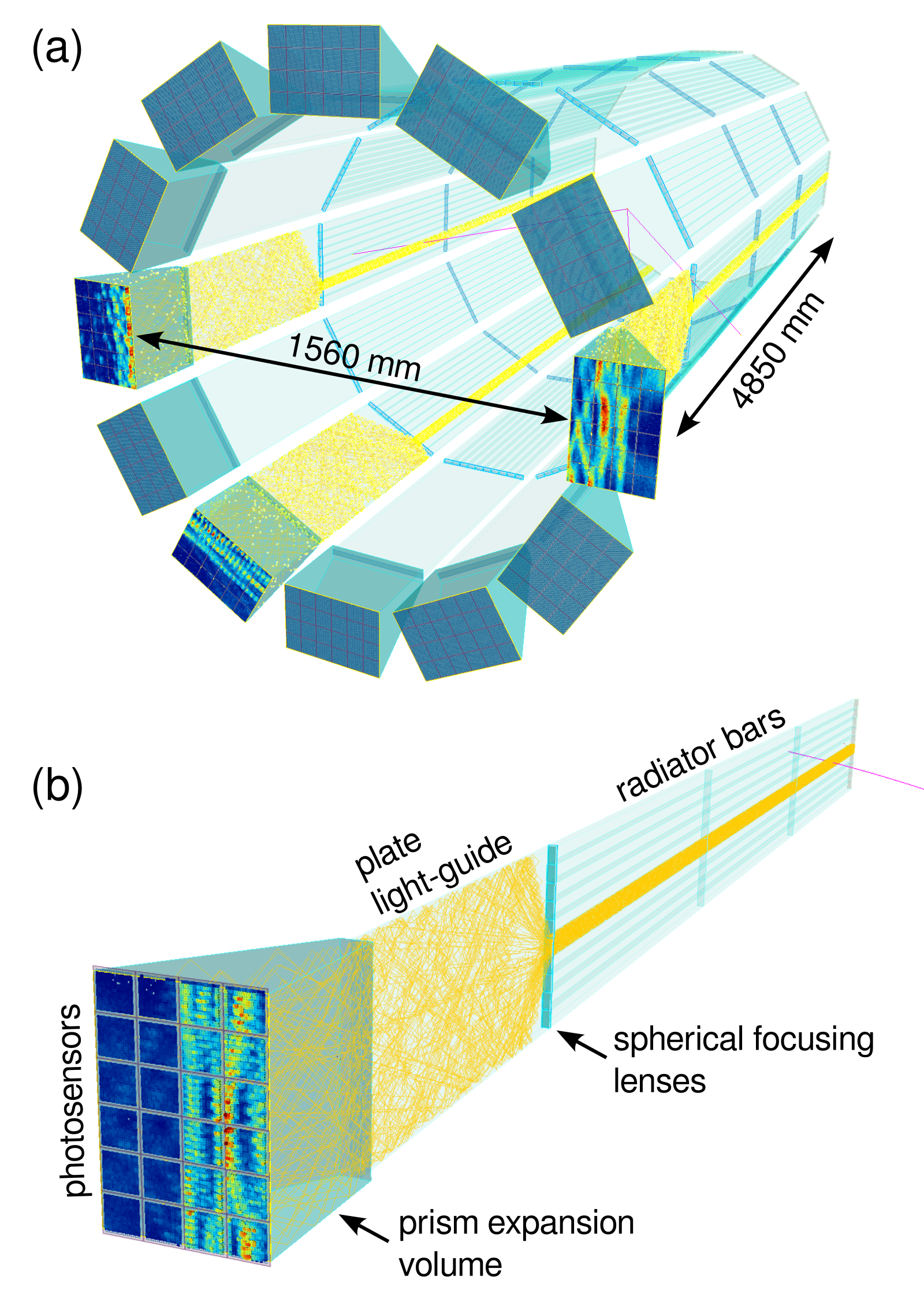}
  \caption{
    Geant4 event display of xpDIRC in barrel configuration with 12 sectors (a)
    and close-up view of an individual sector with highlighted key components (b).
    Cherenkov photons generated by pions are displayed in yellow.
    Hit patterns are accumulated from 5000 simulated tracks.
  }
  \label{xpdirc-geo}
\end{figure}

Fig.~\ref{xpdirc-geo} shows an example of the xpDIRC geometry implemented
in the form of a barrel with 12 isolated sectors (a)
and highlighted key components (b).
The optimization of the design, especially the location and parameters
of the focusing system, was performed using a simulation package based on Geant4 \cite{eicdirc}.
The simulation incorporates realistic wavelength-dependent optical material properties
and was calibrated using experimentally determined values
for relevant parameters such as quantum, collection efficiency,
gain uniformity, PMT's timing resolution of 100 ps, and photon transport efficiency \cite{dirc_tdr}.

\begin{figure}[ht]
  \centering
  \includegraphics[width=0.9\columnwidth]{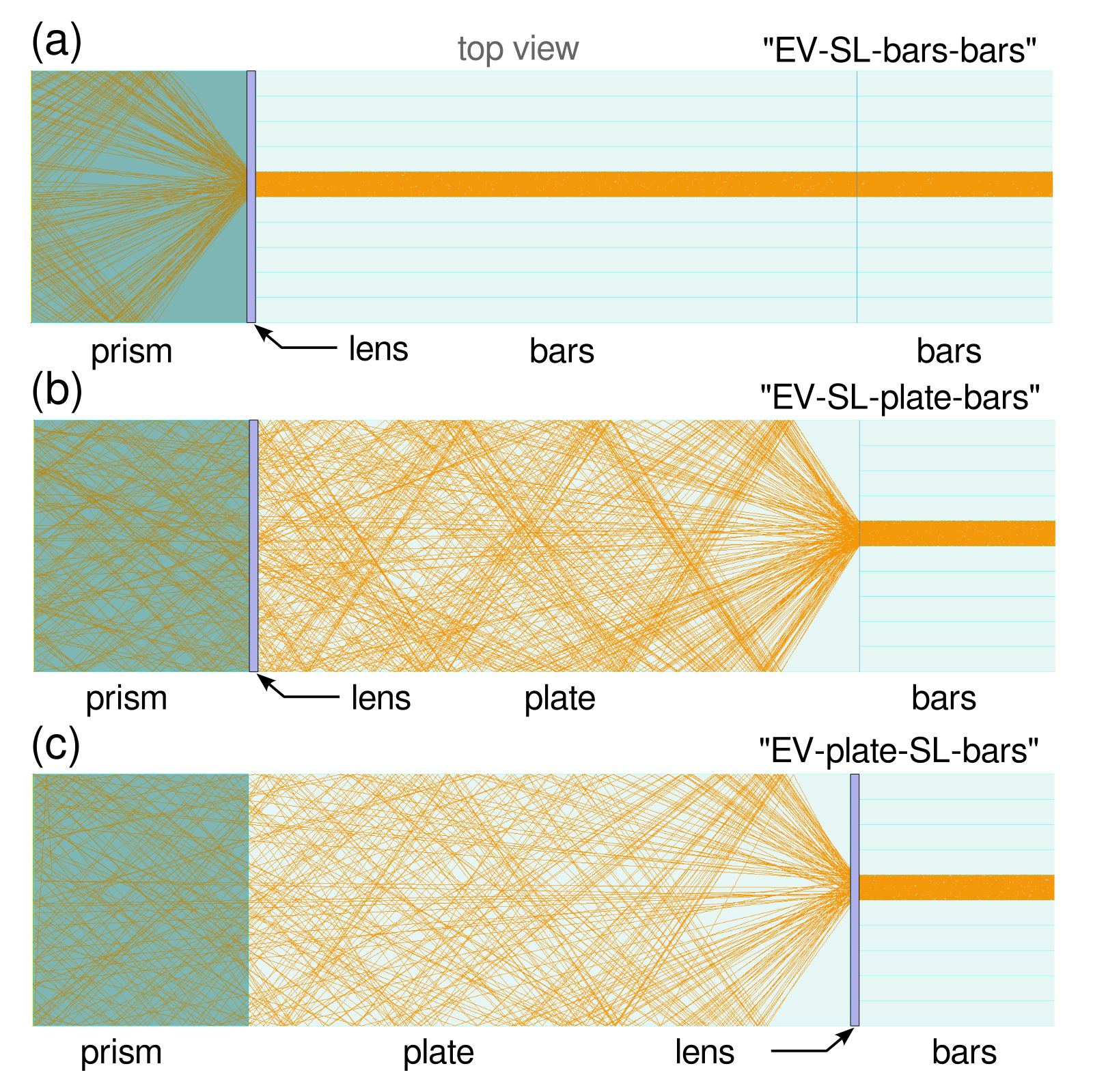}
  \caption{
    Geant4 visualization (top view) of DIRC light guide configurations.
    The paths of Cherenkov photon from a single pion at 6 GeV/c and $30^\circ$ polar
    angle are shown in orange. (a) hpDIRC design using narrow bars for both the active area
    and the light guide section, in combination with spherical lenses (SLs), placed between the
    bars and the prism (EV); (b) conservative xpDIRC hybrid design with narrow bars coupled
    to a wide plate as light guide via a SLs; (c) novel xpDIRC hybrid design with
    the SLs placed between the narrow bars and the wide plate.
  }
  \label{xpdirc-ev}
\end{figure}

The primary approach for enhancing xpDIRC performance centers
on optimizing its optical geometry.
The use of a light guide plate enables longer focal
lengths and improved focusing, which minimizes aberrations and
reduces the required sensor area.
Several optical configurations have been investigated,
with particular attention to lens placement and details of its construction.
Fig.~\ref{xpdirc-ev} shows the Geant4 geometry of three different
configurations - the hpDIRC design (a), a "conservative xpDIRC" 
design (b) and a "novel xpDIRC" design (c).  Also shown in Fig.~\ref{xpdirc-ev}
are the paths of Cherenkov photons from a single pion at 6~GeV/c 
and $30^\circ$ polar angle.
Option (a) shows the design used as a baseline for the ePIC hpDIRC,
with spherical lenses placed between the radiator bars and the EV,
without a plate light guide.
In the conservative xpDIRC design (b) the bars are coupled to the plate
and 3-layer spherical lenses are placed between the plate and the prism.
The novel xpDIRC geometry (c) places the 3-layer spherical lens between
the bars and the plate.
In the latter case, the focal length of the 3-layer spherical lens
was adjusted for the longer distance to the sensors.

\begin{figure}[ht]
  \centering
  \includegraphics[width=0.9\columnwidth]{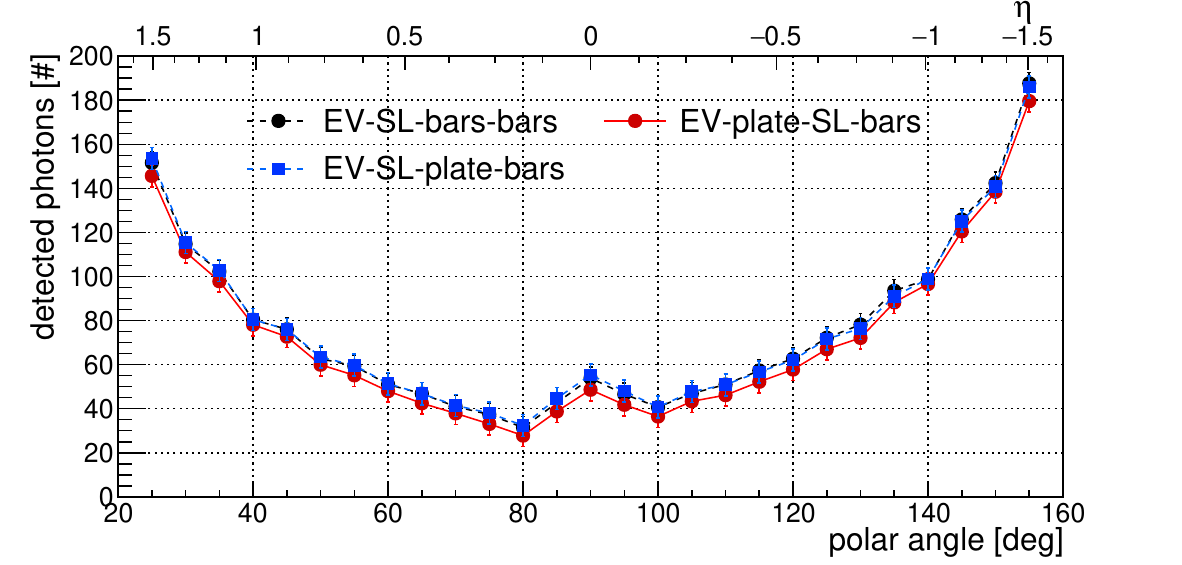}
  \includegraphics[width=0.9\columnwidth]{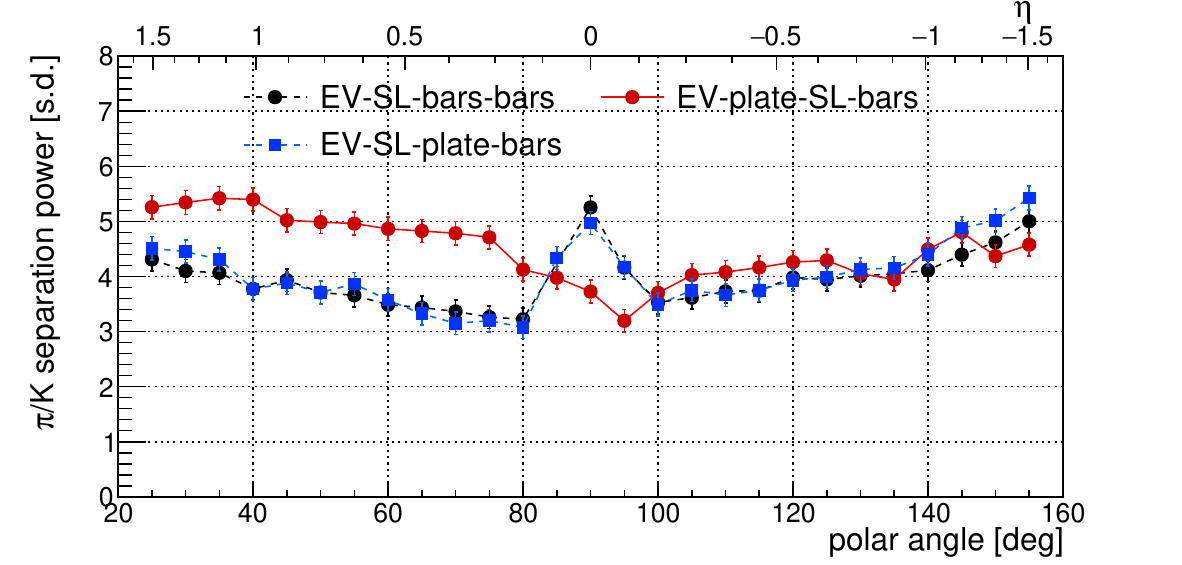}
  
  \caption{
    Photon yield (top) and $\pi/K$ separation power (bottom)
    as a function of the polar angle at 6 GeV/c for different
    DIRC light guide and focusing configurations.
  }
  \label{xpdirc-perf}
\end{figure}

The performance of the configurations was benchmarked using photon yield and $\pi/K$
separation power obtained using the \textit{time imaging} reconstruction method
with simulated probability density functions \cite{tireco}.
The results are shown in Fig.~\ref{xpdirc-perf} as a function of
polar angle and pseudorapidity for 6 GeV/c particles.
The configuration with the focusing lens positioned in front
of the expansion volume showed slightly higher photon yield,
as the lens in this configuration was not subject to strict size
limitations compared to the configurations in front of the light guide plate,
where the lens height was reduced from the original 50~mm to 17~mm
to achieve a slim bar box design with uniform thickness.
The larger height of the lens reduces reflection losses on its sides.

The corresponding $\pi/K$ separation results show
that 3 $\sigma$ is reached for the whole phase space 
and that the novel xpDIRC
configuration with the 3-layer spherical lenses placed
in front of the light guide plate (red)
significantly outperforms the other configurations,
including the baseline performance of the ePIC hpDIRC (black),
in the forward (positive $\eta$) region.

\begin{figure}[ht]
  \centering
  \includegraphics[width=0.9\columnwidth]{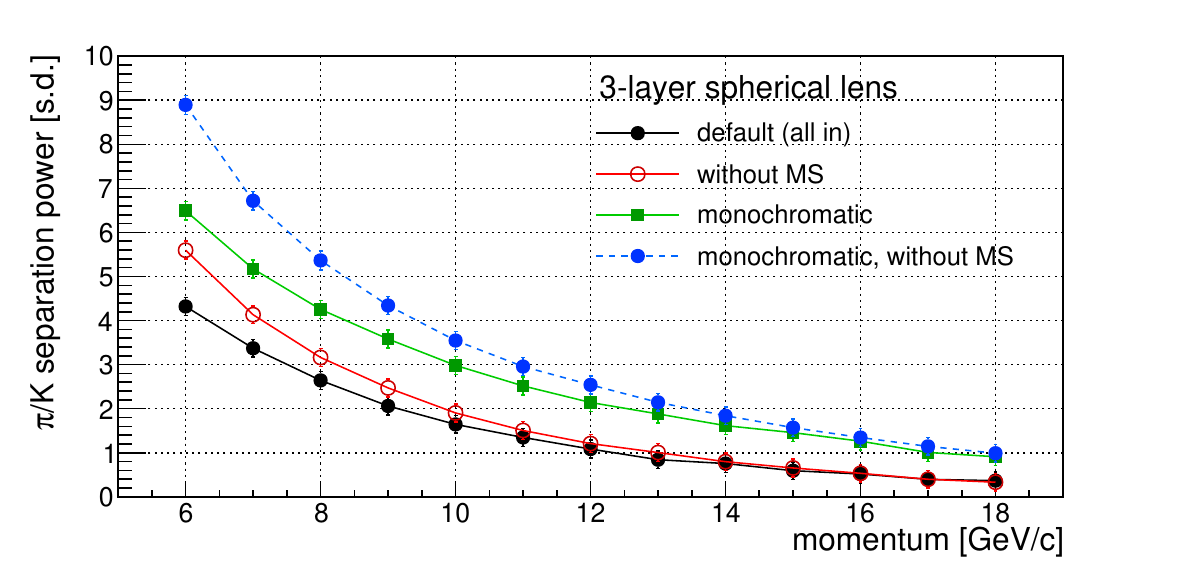}
  \includegraphics[width=0.9\columnwidth]{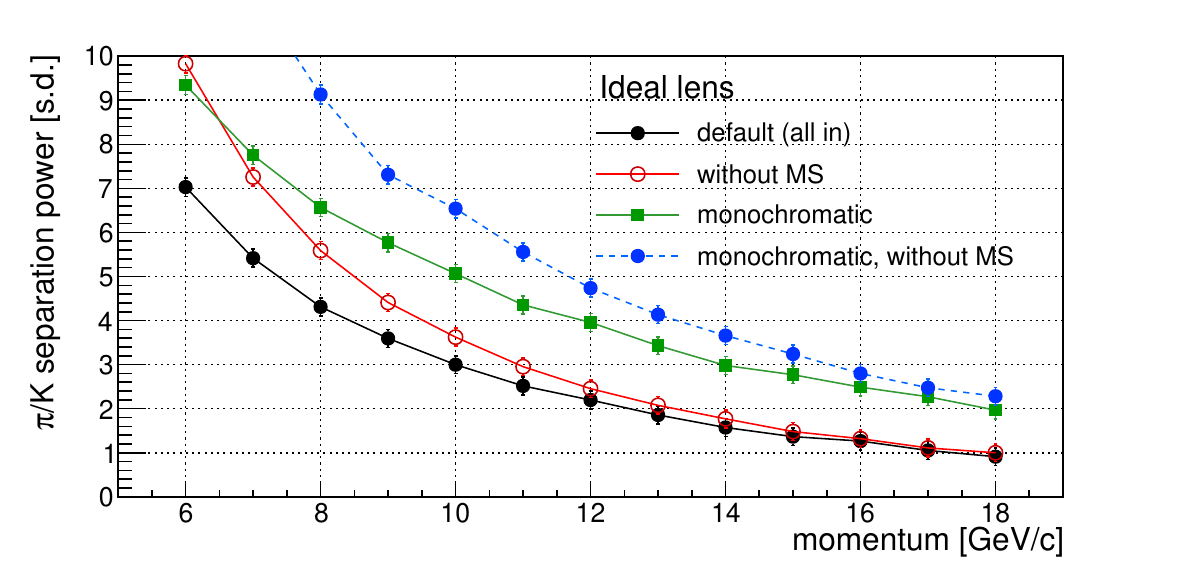}
  
  \caption{ $\pi/K$  separation power as a function of the particle momentum at 30$^\circ$ polar
    angle for the ePIC hpDIRC with two focusing scenarios: 3-layer spherical lens (top) and
    hypothetical ideal focusing without aberrations (bottom).
    Black markers denote the standard Geant simulation process ("default");
    multiple scattering was disabled for the red markers ("without MS");
    chromatic effects in photon emission and propagation were disabled
    by setting the photon energy to 3.2 eV for the green points ("monochromatic");
    both multiple scattering and chromatic effects were disabled for the
    blue symbols.}
  \label{xpdirc-scan}
\end{figure}

\section{Observations}

The promising results for the narrow-bar geometry coupled to a
light guide plate via spherical lenses warranted further investigation.
A follow-up simulation study demonstrated that multiple scattering within the
DIRC bars, as well as optical aberrations in the lenses,
have a significant impact on the DIRC performance.
The ePIC hpDIRC design, with 3-layer spherical lenses
placed between narrow bars and the 30 cm-deep expansion volume prism,
was compared to a hypothetical configuration in which
the lenses were replaced by a mathematical algorithm
performing ideal focusing, without any aberrations.
Fig.~\ref{xpdirc-scan} shows the resulting $\pi/K$ separation
as a function of momentum for different physics processes
enabled or disabled for a standard ePIC hpDIRC configuration (top)
and for the configuration with ideal lens (bottom).
Significantly better performance of the ideal lens
suggests that it may be possible to further improve
the focusing system by reducing aberrations,
for instance, by using aspherical lenses.

The results also suggest that the performance,
particularly in the lower momentum region,
can be improved by mitigating multiple scattering effects.
This can be accomplished by reducing the thickness
of the radiator bars as well as by adding additional
tracking layers located radially outside the radiator.

Further performance improvement can be achieved
by minimizing chromatic effects in the bars.
This can be done by reducing
the optical bandwidth of the DIRC sensors, for instance,
by employing SiPMs or MCP-PMTs with quantum efficiency
shifted toward longer wavelengths \cite{pmts20}.

The xpDIRC hybrid configuration,
with lenses positioned between the bars and the plate,
offers a significant advantage in ring image expansion.
The light guide plate increases the effective expansion distance,
reducing the required depth of the expansion volume. 
A smaller EV not only reduces the material budget
in the endcap region of the detector and simplifies
the integration of the DIRC into the experiment,
but also shrinks the size of the sensor area,
lowering the cost of sensors and electronics
and increasing the signal-to-noise ratio
of the Cherenkov photons relative to the background.
This could be a major advantage in a scenario
where SiPMs are used as photodetectors.

The SiPMs photodetectors offer compelling advantages for Cherenkov detectors,
including low cost, fast timing,
and high photon detection efficiency.
These benefits come with significant challenges.
The primary obstacle is radiation-induced dark count rates (DCR),
which can reach problematic levels.
However, simulations of the ePIC hpDIRC configuration indicate
that even with DCR values up to 100~kHz/mm$^2$ acceptable PID
performance is preserved across most of the angular acceptance.
These results motivate continued R\&D into
SiPM-specific readout strategies, including modeling
additional features such as afterpulsing.
With a more realistic SiPM-based readout model,
we will explore various designs of the expansion
volume to determine the optimal detector plane
size. Once the optimal geometry is identified,
we plan to fabricate a prototype expansion volume,
acquire a new lens, and prepare a matching radiator plate.
These components will be used to test
the xpDIRC concept using the Cosmic Ray Telescope
(CRT) facility at Stony Brook University.

\section{Conclusion}

The xpDIRC represents a forward-looking upgrade
path for DIRC technology, with a design philosophy
focused on enhanced optical performance,
cost-effectiveness, and compatibility with low-power sensor technologies.
A coordinated development effort,
aligned with the ePIC hpDIRC infrastructure
and CRT-based prototype testing, will guide
the next phase of R\&D in the coming years.

\section{Acknowledgements}

This work is supported by DOE EIC-related R\&D program managed
by the Thomas Jefferson Laboratory and by DOE funding DE-FG02-96ER40960.

\end{document}